\documentclass[aoas,preprint]{imsart}
\usepackage[authoryear]{natbib} 
\usepackage{times,mathptmx}
\usepackage{balance} 
\usepackage{amsmath,amssymb,graphicx,amsbsy} %eps figures can be used instead
\usepackage{lastpage}
\usepackage[format=plain,justification=raggedright,singlelinecheck=false,font=small,labelfont=bf,labelsep=space]{caption} 
\usepackage{fancyhdr}
\usepackage{graphicx}
\usepackage{hyperref}
\usepackage{subfig}
\usepackage{color}
\usepackage{paralist}
\usepackage{listings}
\usepackage{xr}
\begin{document}
\begin{frontmatter}

\title{Unusual structures inherent in point pattern data predict colon cancer patient survival}
\runtitle{Point process models predict colon cancer patient survival}

\begin{aug}
\author{\fnms{Charlotte M} \snm{Jones-Todd}\thanksref{t2,a1}\ead[label=e1]{cmjt@st-andrews.ac.uk}},
\author{\fnms{Peter} \snm{Caie}\thanksref{t3,a3}\ead[label = e2]{pdc5@st-andrews.ac.uk}},
\author{\fnms{Janine} \snm{Illian}\thanksref{a1}},
\author{\fnms{Ben C} \snm{Stevenson}\thanksref{a1,au}},
\author{\fnms{Anne} \snm{Savage}\thanksref{a4}},
\author{\fnms{David J} \snm{Harrison}\thanksref{a3}},
\and
\author{\fnms{James L} \snm{Bown}\thanksref{a2}}

\thankstext{t2}{Corresponding author: Charlotte M Jones-Todd, Centre for Research into Ecological $\&$ Environmental Modelling, University of St Andrews, KY16 9LZ, \printead{e1}}
\thankstext{t3}{Corresponding author: Peter Caie, School of Medicine, University of St Andrews, KY16 9TF, \printead{e2}}

\affiliation{\thanksmark{a1}CREEM, School of Mathematics and Statistics, University of St Andrews \thanksmark{a3}School of Medicine, University of St Andrews \thanksmark{au}Department of Statistics, University of Auckland \thanksmark{a2}School of Arts, Media and Computer Games, Abertay University and \thanksmark{a4}School of Science, Engineering and Technology, Abertay University}

\runauthor{Jones-Todd, M, C et.al}
\end{aug}

\begin{abstract}
Cancer patient diagnosis and prognosis is informed by assessment of morphological
properties observed in patient tissue. Pathologists normally carry out this assessment, yet
advances in computational image analysis provide opportunities for quantitative
assessment of tissue. A key aspect of that quantitative assessment is the development of
algorithms able to link image data to patient survival. Here, we develop a point process
methodology able to describe patterns in cell distribution within cancerous tissue
samples. In particular, we consider the Palm intensities of two Neyman Scott point
processes, and a void process developed herein to reflect the spatial patterning of the
cells. An approximate-likelihood technique is taken in order to fit point process models to
patient data and the predictive performance of each model is determined. We demonstrate
that based solely on the spatial arrangement of cells we are able to predict patient
survival.

\end{abstract}

\end{frontmatter}

\section{Introduction}\label{sec:intro}

A fundamental aspect of cancer patient diagnosis and prognosis concerns the assessment by pathologists of morphological properties of  patient tissue sections. These sections typically comprise both cancerous, and non cancerous tissue structures, with regions of each intermixed in space \citep{mattfeldt2014characterization}. Pathologists categorise tumour into stages which are associated with the progression of the cancer and patient outcome (i.e., survival). Cancer staging is good at predicting population survival statistics but not as accurate at predicting an individual patient’s prognosis \citep{caie2016next}. This is due, in part, to a lack of biomarkers within the tissue which can be reliably quantified by eye. The pattern of invasive growth in colon rectal cancer (CRC) has been previously linked to aggressive disease and patient prognosis \citep{pinheiro2014tumor,tokodai2016risk,morikawa2012prognostic}. The phenomenon of tumour budding, where small distinct islands of tumour cells, widely dispersed within the stroma at the invasive front of CRC, has been shown to be prognostically significant \citep{rieger2017comprehensive,de2016tumour}. Advances in computational image analysis provide an opportunity for quantitative, and objective assessment of biomarkers and tissue morphology. \citet{caie2014quantification} have analysed colorectal cancer tissue sections in such a fashion. Although they measured the morphological pattern of the tumour they did not take into consideration their spatial arrangement. 

\citet{mattfeldt2014characterization} explore the use of spatial patterning of cells as an indicator of patient outcome and use point process methodology to characterise tissue structures. They compare the estimated point process intensity, the second-order statistics pair correlation function, and Ripley's K-function between two patient groups. They compared patients whose cancer had and had not metastasised, a major indicator of patient survival. Results, however, showed no differences between patient groups with respect to first- or second-order spatial statistics; the authors recognise that standard statistical point process methods are not appropriate in capturing the morphology of tissue structure arising from the spatial intermixing of tumour and stroma. As such, we propose a novel approach novel approach using nonstandard point processes. This through utilising concepts from spatial statistics, specifically point process methodology, in order to predict patient survival solely based on the spatial arrangement of cells.

In general terms, point pattern data \citep{illian2008statistical,gelfand2010handbook,moller2007modern,diggle2013statistical} describe the location of events or objects (such as cells), typically in two-dimensional space. Such data may exhibit different structural properties, often classified as spatial randomness, clustering, or regularity. 
Point pattern data are realisations of point processes, and perhaps the most discussed in the literature is the Poisson process. A Poisson process is a process in which the number of points follows a Poisson distribution, and the points are independently scattered. The homogeneous Poisson process can be thought of as a basic point process, as it describes one of the most simple of point pattern structures: complete spatial randomness (CSR). Notwithstanding the appeal of this simple characterisation, rarely in the physical world are point patterns homogeneous. For example, cluster point processes are used by \citet{illian2013fitting} in considering the locations of muskoxen \textit{Ovibos moschatus} herds in Zackenberg valley, and \citet{stoyan2000recent} in the context of forestry data. In addition to cluster processes, but perhaps considered to a lesser extent, void (areas devoid of any points) dynamics are also inherent in the natural world. For example with forestry data, understanding the void dynamics aids in inferring the local interaction among tree species \citep{dube2001quantifying}. Voids are also inherent in astrological data, and an algorithmic approach is often used in their detection \citep{kauffmann1991voids}.

While tissue sections exist at a much smaller scale, such structures are still implicit. The morphological patterns within tissue sections, at the very least, \begin{inparaenum}[(i)] \item are non-homogeneous, and \item exhibit abstruse spatial morphology\end{inparaenum}. In this article we propose the use of point process models to quantify the complex spatial structures in tumour tissue sections. In particular, we use the interpoint distances between cells to inform consideration of the spatial morphology of the tissue. We follow the estimation methodology initially proposed by \citet{tanaka2008parameter} to fit a Thomas cluster process---a type of Neyman Scott point process---and extend this to estimation of parameters of a Mat\'{e}rn cluster process, and the void process derived herein. Section \ref{sec:processes} discusses the morphology of the above mentioned processes, introducing the Palm intensity function. Section \ref{sec:method} describes the approximate-likelihood framework proposed by \citet{tanaka2008parameter}, which is then generalised to incorporate estimation of parameters of both the void and Mat\'{e}rn cluster processes. Section \ref{sec:cancer} presents the results of the application of our approach to a colorectal cancer patient data set as described in \citet{caie2016next} and \citet{caie2014quantification}.

\section{Cluster and void processes}\label{sec:processes}

Generally and across many problem domains, point pattern data exhibit a diverse range of configurations \citep{penttinen2007statistical,soubeyrand2014nonstationary}, including mixtures of clusters of points and regions devoid of points where one might expect points to occur given the structure of the overall pattern. Cluster and void processes are used to model data that exhibit the structures referred to by their names respectively. Yet, when considering real-world phenomena, it may not be straightforward to identify which process pertains to the data. In the case of cancerous tissue morphology, one may think that areas of necrosis constitute voids, or that tumour cells constitute clusters. In reality, the morphology of cancerous tissue is a result of many complex processes, resulting in obscure spatial structures capturing the intermixing of cancerous and non cancerous cells. This section proceeds to introduce both a cluster and a void process, discussing their relevance in capturing complexities inherent in point pattern data. That may be non-homogeneity due to spatial clustering  of cells, spatial intermixing of tumour and stroma, or regions devoid of any particular cell, caused perhaps by necrosis. Thus, characterising the spatial structure of cells is nontrivial, and requires the use of more complex, nonstandard point process methods.

Consider a Neyman Scott point processes (NSPP); a type of point process which gives rise to clustered point patterns. These clusters of points are assumed to be the result of unobserved parent points who sire a number of observed daughters. The obstacle is in estimating the number of parents, having only observed the daughters. Here, we propose that void processes can be constructed in a similar hierarchical fashion. That is, there are assumed to be unobserved parents which dictate the structure of the observed pattern, in this case resulting in voids rather than clusters.

In each case there exist unobserved parent points, which themselves are realisations of a point process with homogeneous intensity. These parent points induce structures in the point pattern, that is, either void spaces or clusters of points. Herein the observed points of the pattern are termed daughters, which are either repelled (for a void process), or sired (for a cluster process), by the unobserved parent points. This concept is illustrated in Figure \ref{fig:simulated}. In each case, the intensity of the unobserved parent points can be thought of in terms of a latent process driving the spatial arrangement of daughters.

\subsection{A void process}\label{subsec:voids}

A void process has unobserved parent points that repel their observed daughters. That is, assuming a homogeneous pattern of points, parents expunge every daughter in a sphere of radius R centered at the parent's location. The remaining daughters are free from their parent's influence, and are thus realisations of a void process. A realisation of a void process is shown in Figure \ref{fig:simulated}, plot $i)$, where it is relatively easy to identify voids by eye. A slightly more robust method than identifying voids by eye---the use of a 4-neighbour graph---is used by \citet[p.~258]{illian2008statistical} in order to identify areas of storm damage in a plot of trees. However, for sparser patterns, and patterns with voids in close proximity, this spatial patterning becomes much more difficult to describe or identify. Following the methodology described below, the distribution of distances between two independent daughter points is used to estimate the parameters.

A void process is defined by both the background intensity of daughters, the density of parents, and the radius of the voids. The parents are a realisation of a Poisson point process with density D. The distribution of the distance between two independent daughters depends on the expunging radius, R, of the parents. In this article a potential daughter is termed to be "safe" if it is not within distance R of any parent, and therefore it is observed. Any region that is not within distance R of any point is called a safe region. The parameters of a void process are then $\boldsymbol{\theta}=(D,R,\lambda)$. Where $\lambda$ is the density of observed daughters (i.e., those daughters outwith the voids). A void process, therefore, is an example of a dependently thinned point process \citep[p.~365]{illian2008statistical}.

Little headway has been made into processes exhibiting void dynamics; importantly, similarities may be made between void processes and repulsive processes. Processes giving rise to repulsive point patterns, such as Gibbs or Strauss processes, are generalisations of the homogeneous Poisson process. In each case the points are no longer independently distributed in space; in fact points are repelled by one another, and this repulsion is constant within a fixed interaction radius $r_0$ around each point (for a Strauss process). The strength of this interaction ranges from no interaction---equivalent to a homogeneous Poisson process---to outright inhibition within the radius $r_0$ around each point---where no points are observed within distance $r_0$ from any observed point.  Along a similar vein, another class of interaction process, a hardcore process, cannot contain points that are closer than a distance, $r_0$, apart. One such process is termed a hardcore Mat\'{e}rn process which is a result of some dependent thinning operation \citep{matern2013spatial}. Here points do not exist---due to the thinning operation---within a certain radius, $r_0$, of each point in the observed (thinned) point process. At first glance, hardcore processes might be misinterpreted as void processes. However, by definition they describe regularity in point patterns, which is due to some inhibitive interaction \citep{stoyan1988thinnings} between all observed points; whereas realisations of void processes are a consequence of unobserved parents and therefore only exhibit areas of inhibition. 

\subsection{Cluster processes}\label{subsec:cluster}

In the context of the cluster processes discussed here, there exist unobserved parent points, which themselves are realisations of a point process with homogeneous intensity. These parent points sire observed daughter points, inducing clustering. The intensity of the unobserved parent points can be thought of in terms of a latent process driving the arrangement of the clusters of daughters. This section defines Neyman Scott point processes as in \citet{illian2008statistical}, and discusses two examples of the Neyman Scott point process---namely, the Mat\'{e}rn cluster process, and the modified Thomas process.

An isotropic Neyman Scott point process is defined by three components: \begin{inparaenum}[(i)] \item the distribution of daughters sired by each (unobserved) parent, \item the distribution of the number of daughters sired by each (unobserved) parent, and \item  the distribution of daughter points around their parents. \end{inparaenum} The latter can also be partitioned into two aspects, those being the distribution of the distances between two independent daughter points sired by \begin{inparaenum}[(i)] \item the same, and \item different parents. \end{inparaenum}

Characterising the distribution of daughters sired by each parent enables the derivation of the distance distributions. In the case of the Mat\'{e}rn cluster process the daughters are uniformally distributed in a circle around their parents, whereas the modified Thomas process has daughters which are bivariate normal around each parent. These subtle differences portend to different model parameters, yet similar estimation procedures as that proposed by \citet{tanaka2008parameter} can be used.
Parameters of a Neyman Scott point process are given by $\boldsymbol{\theta} = (D,\phi,\gamma)$. Here, $D$ is the intensity of the unobserved parent points, which are assumed to be realisations of a homogeneous Poisson process. The number of daughters sired by each parent are assumed to be IID from some discrete distribution characterised by the parameter $\phi$. In the case of a Poisson distribution $\phi$ would be the expectation. Conditional on their parents' locations, the daughters are scattered in space according to some distribution with parameter $\gamma$. %% In the case of a Mat\'{e}rn this corresponds to the radius of the circle within which the daughters are scattered uniformally. In the case of the modified Thomas process $\gamma$ corresponds to the Gaussian dispersion of daughters around their parents.

\begin{figure*}[htp]

\centering
\includegraphics[width=.3\textwidth]{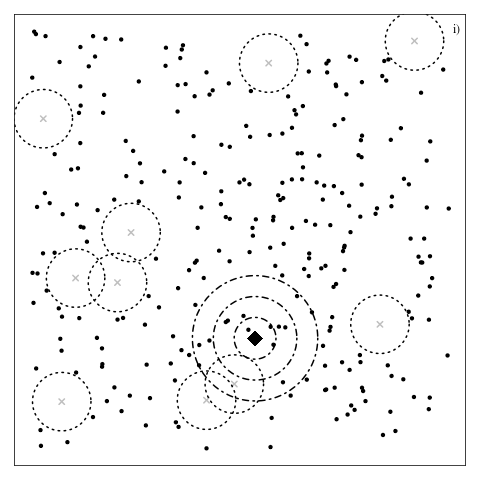}
\includegraphics[width=.3\textwidth]{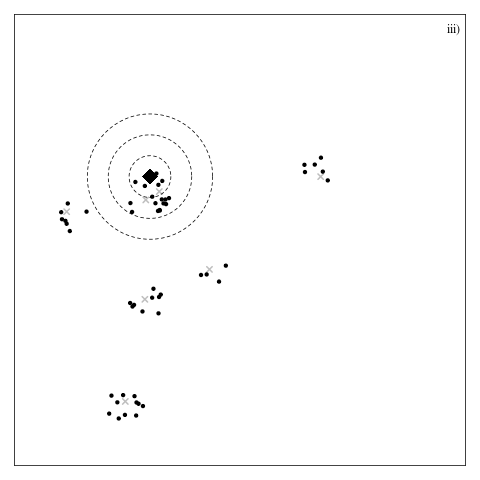}
\includegraphics[width=.3\textwidth]{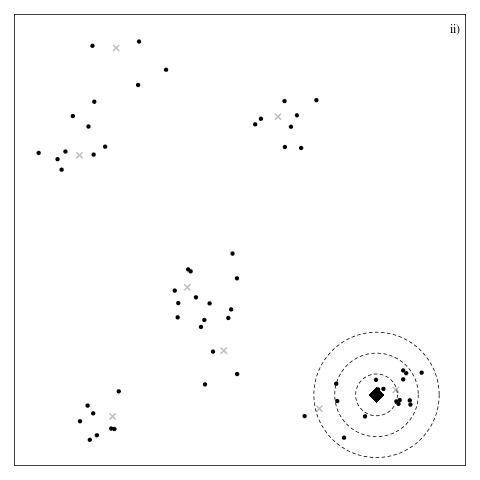}

\caption{i) A simulated void processes in the unit square. With (unobserved) parent points (crosses) expunging daughters within a distance $R$ so that only the observed daughters remain (dots). The daughters are generated by a homogeneous Poisson process with densities $\lambda=300$. The number of parents simulated are IID following a Poisson distribution with expectation $D=10$. The distance beyond which daughters are safe from their parents is $R=0.075$. The dotted circles indicate the simulated voids. Noting an arbitrarily chosen daughter (encircled diamond) the distance between her and her sibling is clearly related to the proximity of a parent. The dashed circles show how the intensity of observed daughters changes for different distances $r$ away from an observed daughter. Plots ii) $\&$  iii): two simulated Neyman Scott point processes in the unit square; Thomas \textit{ii)}), and Mat\'{e}rn \textit{iii)}), with unobserved parent points (crosses) siring the observed daughters (dots). In each case parents are generated by a homogeneous Poisson process with intensity $D = 7$. The number of daughters sired by each parent are IID following a Poisson distribution with expectation $\phi = 8$. In each case $\gamma = 0.05$. In the left hand plot daughters are dispersed around their parents due to a bivariate normal distribution, $\mathbf{N_2}(\mathbf{0},\boldsymbol{\sigma^2}\mathbf{I_2})$, where $\gamma = \sigma = 0.05$, in the right hand plot daughters are uniformally distributed in a sphere with radius $R$ around their parents, hence $\gamma = R =0.05$. }
\label{fig:simulated}

\end{figure*}

Figure \ref{fig:simulated} (plots \textit{ii)} and \textit{iii)}) show realisations of a Thomas and Mat\'{e}rn process respectively. Given daughter locations alone (dots) the data may be considered to be realisations of Neyman Scott point processes with lower parent density, and larger $\phi$ and $\gamma$. Here the data have been chosen to illustrate this problem as they are aggregated such that two parents in each case are close-by, hence attributing cluster identity to daughters is unclear.

\section{Methodology}\label{sec:method}

The approach taken here in order to estimate the parameter $\boldsymbol{\theta}$ for both the void and the cluster process discussed in Section \ref{sec:processes} is based upon the derivation of the Palm intensity function. The Palm intensity function in each case, is parameterised by the vector $\boldsymbol{\theta}$, which reflects the structure of each process. Details of each Palm intensity (i.e., void and cluster process), are detailed in Sections \ref{sec:voidpalm} and \ref{sec:nspp} respectively. Derivation of these Palm intensities in d-dimensions is given in Appendix A. Our application requires only the Palm intensities in 2-dimensions, hence, only the 2-dimensional cases are discussed in the main body of this article. 

\subsection{Maximum Palm likelihood estimation}

The likelihood for a Neyman-Scott point process is widely held to be intractable \citep{waagepetersen2007estimating}. However, an approach taken by \citet{tanaka2008parameter} shows that the pairwise distances can be approximated by a non-homogeneous Poisson point process, the intensity of which being proportional to the Palm intensity function. The likelihood for this approximating model is termed the Palm likelihood. \citet{prokevsova2013asymptotic} show that such Palm likelihood estimators are consistent. In addition to using a Palm likelihood approach to estimate parameters of the Thomas and Mat\'{e}rn processes (Neyman Scott point processes) we also take such an approach to estimate the parameters of the void process.

\subsubsection{The Palm intensity function}\label{sec:palm}\hspace{0.5cm}
The Palm intensity, $\lambda(r;\boldsymbol{\theta})$, describes how the point density varies as a function of distance $r$ from an arbitrarily chosen point. Thus, the Palm intensity function returns the expected intensity of a point process at a distance $r$ from an arbitrarily chosen point. Figure \ref{fig:simulated} illustrates how the point density in a circle of radius $r$ centered at an arbitrary observed point, $b(x,r)$ (dashed circles), changes as $r$ increases. That is, at short distances the presence of a point suggests that there may exist a nearby sister. This is true for both the void and cluster processes, as, in the former case, conditional on observing a daughter the closest a possible parent can be is $R$. Therefore, any potential point is in danger if a  parent exists in the circles of radius $R$ centered at the potential point minus the intersection of the two circles (for details see Appendix). In the latter case, by definition a clustered pattern infers that conditional on observing a daughter there is a high chance that another exists nearby. 

The expected intensity of the point process at a distance $r$ from an arbitrarily chosen point depends on the parameters described in Section \ref{sec:processes} above. The Palm intensity, $\lambda(r;\boldsymbol{\theta})$, can be defined as a function of the parameters $\theta$ that changes over distance $r$, decaying to the overall average intensity of the point process.

The following subsections derive the Palm intensity for a void process (Section \ref{sec:voidpalm}), details the derivation of the Palm intensity for the Thomas process as \citet{tanaka2008parameter}, and generalises this to the Mat\'{e}rn Palm intensity (Section \ref{sec:nspp}).

\subsubsection{Void process}\label{sec:voidpalm} 

Figure \ref{fig:simulated}, plot \textit{i)}, illustrates that considering an observed daughter (the diamond), a potential nearby point is also likely to be safe. The Palm intensity of a void process is the product of the global point density of the pattern and the probability that a potential point at distance $r$ is in a safe area. The aforementioned probability is related to the geometry of the intersection between circles of common radius $R$ centered at the observed daughter (see Figure \ref{fig:simulated} plot \textit{i)}, encircled triangle) and a potential point. This is due to the expunging distance of parents, see Figure \ref{fig:simulated}, plot \textit{i)} dotted circles. Further details are given in Appendix A, and illustrated  therein (see Figure \ref{fig:intersection}).

The area of intersection between two 2-dimensional circles with a common radius R is given by $I(r)$, where,
\begin{equation}\label{eq:intersection}
  \begin{aligned}
    I(r) & = \pi \, R^2\,\text{I}\left(1-\left(\frac{r}{2R}\,\right)^2;\frac{3}{2},\frac{1}{2}\right).\\
  \end{aligned}
  \end{equation}
Here $\text{I}(z;a,b)=\frac{B(z;a,b)}{B(a,b)}$ is the regularised Beta function. 
The Palm intensity function is then derived as (see Appendix A),
\begin{equation}\label{eq:void.palm}
  \begin{aligned}
\lambda(r;\boldsymbol{\theta}) & =\lambda\,\text{exp}\left(-D\,\pi \, R^2\,\left[1-\text{I}\left(1-\left(\frac{r}{2R}\,\right)^2;\frac{3}{2},\frac{1}{2}\right)\right]\right),\\
& = \lambda\,\text{exp}\left(-D\,\pi \, R^2\,\left[1-\text{F}_{g(r)}\left(\frac{3}{2},\frac{1}{2}\right)\right]\right),
\end{aligned}
\end{equation}
  where $g(r)=1-\left(\frac{r}{2R}\,\right)^2$, and $\text{F}_{g(r)}(\cdot,\cdot)$ is the CDF of the Beta distribution. Note when $r=0$ $\Rightarrow$ $g(r)=1$ $\Rightarrow$ $\text{F}_1(\cdot,\cdot)=1$ $\Rightarrow$ $\lambda(0;\theta)=\lambda$, in addition when $r=2R$ $\Rightarrow$ $g(r)=0$ $\Rightarrow$ $\text{F}_0(\cdot,\cdot)=0$ $\Rightarrow$ $\lambda(0;\theta)=\lambda\,\text{exp}(-D\,\pi \, R^2)$, due to the properties of the CDF.
  
  Figure \ref{fig:palms}, plot \textit{i)}, shows both the empirical Palm intensity function (solid line), and the fitted Palm intensity (dashed line) for the simulated void process shown in Figure \ref{fig:simulated}. The horizontal asymptote to which the Palm intensity decays is the global point density, the Palm intensity is a piece-wise continuous function with two sub-domains $(0,2R],[2R,\infty)$.
  
\subsubsection{Neyman Scott point process}\label{sec:nspp} 

Under the conditions that \begin{inparaenum}[(i)] \item parent locations are realisations of a homogeneous Poisson process, \item the number of daughters sired by each parent are Poisson IID, and \item are dispersed due to a bivariate normal distribution with a variance-covariance matrix of any (positive-definite) form, \end{inparaenum} \citet{tanaka2008parameter} derived the analytic Palm intensity for a modified Thomas process, as, 
  \begin{equation}\label{eq:tanaka.palm}
\lambda(r;\boldsymbol{\theta})= D\, \nu + \frac{\nu}{4 \pi \sigma^2}\text{exp}\left(\frac{-r^2}{4 \sigma^2}\right),
  \end{equation}
where $D\,\nu$ is the global point density. Thus, $\lambda(r;\boldsymbol{\theta})$ is the sum of the intensity of non-sibling points, $D\,\nu$, and a Gaussian function describing the intensity due to sibling points. 

This Palm intensity function (Equation \ref{eq:tanaka.palm}) has been extended to the d-dimensional case by \citet{stevensonphd}. Yet, as our application is in 2-dimensions we only discuss the 2-dimensional case here; thus, in general Equation \eqref{eq:tanaka.palm} is given by,
\begin{equation}\label{eq:palm.intensity}
  \begin{aligned}
  \lambda(r;\boldsymbol{\theta}) &= D\,\nu + \frac{\nu\,f_y(r;\gamma)}{2\,\pi\,r}.
  \end{aligned}
  \end{equation}
  
The parameter $D$ is as discussed in Section \ref{sec:nspp} above. Letting $Y$ denote the distance between two randomly selected siblings (i.e., daughters sired by the same parent). Then $f_y(r;\gamma)$ is the PDF of this random variable, which is characterised by the parameter $\gamma$ pertaining to the form of distribution of daughters around their parents.

This PDF of the random variable, $Y$, refers to, in the case of the modified Thomas process, the distance between two normally distributed siblings. For example, in Equation \ref{eq:tanaka.palm},
\begin{equation}\label{eq:thomas.dist}
  f_y(r;\gamma) = \frac{r\:\text{exp}(-r^2/(4 \gamma^2))}{2\:\gamma^2}, 
  \end{equation}
where, $\gamma = \sigma$, is the parameter describing the Gaussian dispersion of daughters around their parents.

\begin{figure}
  \centering
  \includegraphics[width=0.45\textwidth,height = 0.3\textheight]{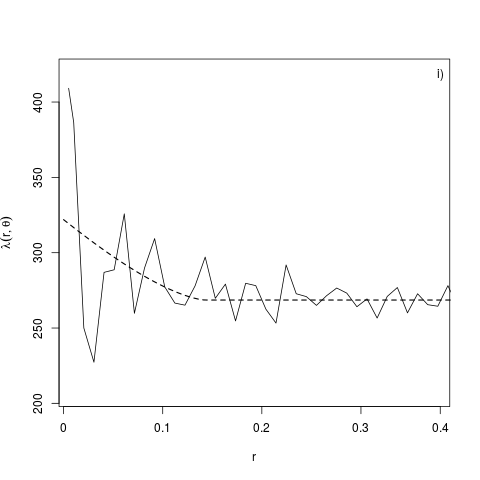}
  \includegraphics[width=0.45\textwidth,height = 0.3\textheight]{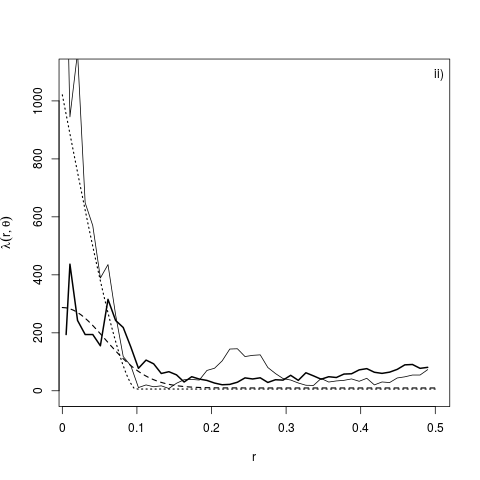}
  \caption{Both the empirical---solid lines---and fitted---dashed lines---Palm intensities for the simulated patterns above (see Figure \ref{fig:simulated}). The Palm intensity for the void process is shown in plot \textit{i)}, it is evident that this function is a piece-wise continuous function with two sub-domains $(0,2R],[2R,\inf)$. The Palm intensities for both the simulated Neyman Scott point processes are shown in plot \textit{ii)}, where the Mat\'{e}rn cluster process (thinner solid line), decays at a much steeper rate, due again to it being a piece-wise continuous function with two sub-domains $(0,2R],[2R,\inf)$. In each case the Palm intensities decay to the global point density of the process.}
  \label{fig:palms}
  \end{figure}

In considering the Mat\'{e}rn process, we follow the same methodology in deriving the Palm intensity function of the aforementioned process. The random variable $Y$, is now the distance between two siblings that are uniformally distributed in a 2-dimensional sphere of radius $R$ around their parents. The PDF of $Y$ is shown below \citep{tu2002random}, using both the hyper-geometric function and integral representation respectively,
\begin{equation}\label{eq:mat.dist}
  \begin{aligned}
    f_y(r;R) &= \frac{4}{B(\frac {3} {2},\frac{1}{2})}\, \frac{r} {R^3}
    \times \left[ {}_2F_1\left( \frac {1} {2},- \frac {1} {2}, \frac {3} {2},1\right)\,R - {}_2F_1\left(\frac {1} {2},- \frac {1} {2}, \frac {3} {2}, \frac {r^2}{4\,R^2}\right)\,\frac {r} {2} \right], \\
    & = \frac {4 \,r\, \int^R_{\frac{r}{2}} (R^2 - x^2)^{\frac {1}{2}} dx} {B(\frac {3} {2},\frac{1}{2})\, R^{4}}.
    \end{aligned}
  \end{equation} 
Here $B(\cdot,\cdot)$ denotes the beta function, and ${}_2F_1(\cdot,\cdot,\cdot,\cdot)$  the hyper-geometric function. 

Figure \ref{fig:palms}, plot \textit{ii)} shows the Palm intensities for the simulated Neyman Scott point processes in Figure \ref{fig:simulated} (the Thomas process, plot \textit{ii)}; Mat\'{e}rn process, plot \textit{iii)}). The solid lines are the empirical palm intensities for each process (Mat\'{e}rn process---thinner line, Thomas process---thicker line). Dashed lines (Mat\'{e}rn process---short dash, Thomas process---long dash) are the respective fitted Palm intensities. Both simulated processes in Figure \ref{fig:simulated}, plots \textit{ii)} and \textit{iii)}, were simulated with the same parameter values $\boldsymbol \theta = (7, 8, 0.05 ) $. Hence, the Palm intensities in each case decay to the same asymptote, the global point density $D\nu$. The Palm intensity of the Mat\'{e}rn process is initially a lot higher, and decays at a much faster rate to that of the Thomas process with the same parameterisation; this as, the parameter $\gamma$ controls the dispersion of daughters around their parents (the radius of dispersion in a Mat\'{e}rn process and the spread in a Thomas process). For the Mat\'{e}rn process $\gamma$ is a hard boundary which induces a much higher intensity at shorter distances. Also of note is the similarity between the Palm intensity of the Mat\'{e}rn process and the void process, namely the shape (piece-wise continuous with two sub-domains $(0,2R],[2R,\infty)$), and this is again due to the geometry of the structure.

\subsection{The Palm likelihood}\label{sec:likelihood}
Taking an approximate-likelihood approach, the estimator for $\boldsymbol{\theta}$ is given by
\begin{equation*}
  \hat{\boldsymbol{\theta}} = \text{arg}\,\text{max}_{\boldsymbol{\theta}}\,L(\boldsymbol{\theta};\mathbf{r}),
  \end{equation*}
which is evaluated through the numerical maximisation of $\text{log}(L(\boldsymbol{\theta};\mathbf{r}))$ with respect to $\boldsymbol{\theta}$. Here, $L(\boldsymbol{\theta};\mathbf{r})$ is termed the Palm likelihood and given by,
\begin{equation}\label{eq:likelihood} 
 L(\boldsymbol{\theta};\mathbf{r}) = \left( \prod n\,\lambda(r;\boldsymbol{\theta}) \right)\,\text{exp}\left(-n \int_0^t \lambda(r;\boldsymbol{\theta})\,2\,\pi\,r\:\text{d}r  \right),
   \end{equation}

where the integral is a volume of $2$ dimensions around the intensity axis. Below, the integral is simplified for each process discussed herein, providing a closed-form Palm likelihood function in each case. This results in an objective function that is very computationally efficient to compute.

\begin{itemize}
\item \textit{Void process}
The integral for the void process is intractable as it contains the CDF of a Beta distribution, however it can be simplified.
\begin{equation*}
    \int_0^t \lambda_o(r;\boldsymbol{\theta})\,2\,\pi\,r\:\text{d}r   = \lambda\,2\,\pi\,\text{exp}\left(-D\,\pi \, r^2\:\right)\,\int_0^t\text{exp}\left(D\,\pi\,r^2\:\text{F}_{g(r)}\left(1,\frac{1}{2}\right)\right)\, r\,\text{d}r.
  \end{equation*}
Refer to Appendix A for a derivation of these.
  
\item \textit{Neyman Scott point processes}
 The integral simplifies as,
\begin{equation*}
 \int_0^t \lambda_o(r;\boldsymbol{\theta})\,2\,\pi\,r\:\text{d}r = D\,\nu + \nu \:\,F_y(t;\gamma),
   \end{equation*}
where $F_y(t;\gamma)$ is the CDF of the distance between two randomly distributed siblings. Below are each $F_y(t;\gamma)$ for each case of the Neyman Scott point process.

\begin{itemize}
\item \textit{Mat\'{e}rn cluster process}
\begin{equation*}\label{eq:cdf.matern}
  F_y(t;R)  = \frac{t^{2}}{R^{2}}\left(1-\frac{B_{\alpha}\left(\frac{1}{2},\frac{3}{2}\right)}{B\left(\frac{1}{2},\frac{3}{2}\right)}\right) + 4\,\frac{B_{\alpha}\left(\frac{3}{2},\frac{3}{2}\right)}{B\left(\frac{1}{2},\frac{3}{2}\right)},
  \end{equation*}
where $B_\alpha(\cdot,\cdot)$ and $B(\cdot,\cdot)$ are the incomplete beta function and the beta function respectively. That is  $B(\alpha;a,b) = \int_0^\alpha u^{(a-1)}(1-u)^{(b-1)} du$, and for $\alpha = 1$ $ B(\alpha;a,b)  = B(a,b)$.  The derivation of these equations is detailed in Appendix A.

\item \textit{Thomas cluster process}
 Letting $q = \sigma_\chi \sqrt{2}$, which is strictly monotonic. Then, showing the derivation of \citet{stevensonphd},
\begin{equation*}
 F_q(q;\sigma) = P\left(1,\frac{4 q^2}{\sigma^2}\right)
\end{equation*}
where $P(\cdot, \cdot)$ is the regularised gamma function.
\end{itemize}

\end{itemize}

\section{Application to histopathology data}\label{sec:cancer}

\begin{figure*}[htp]

\centering
\includegraphics[width=.3\textwidth]{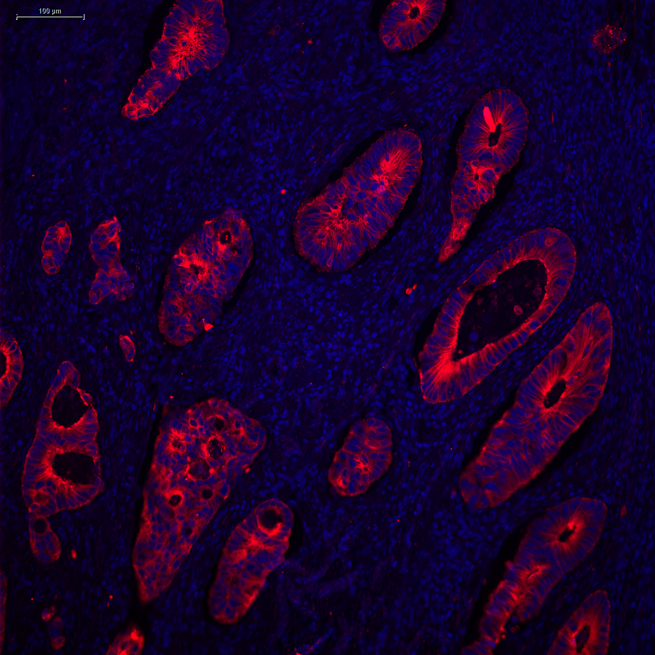}
\includegraphics[width=.3\textwidth]{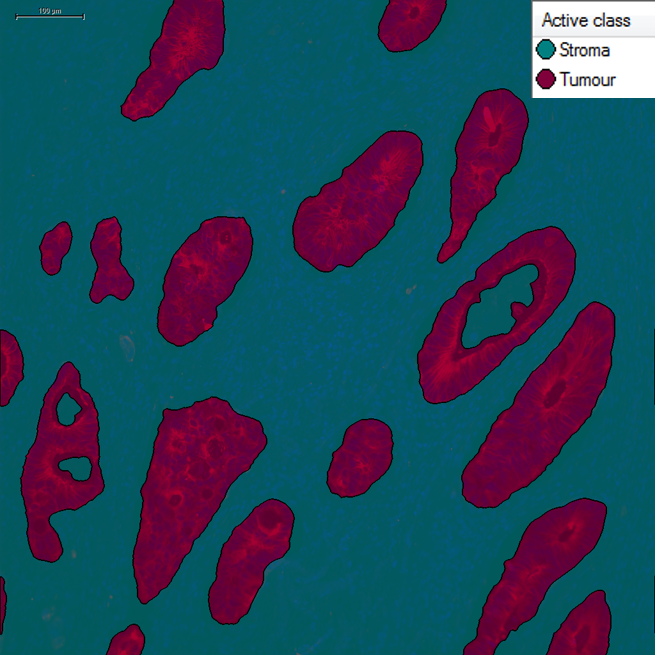}
\includegraphics[width=.3\textwidth]{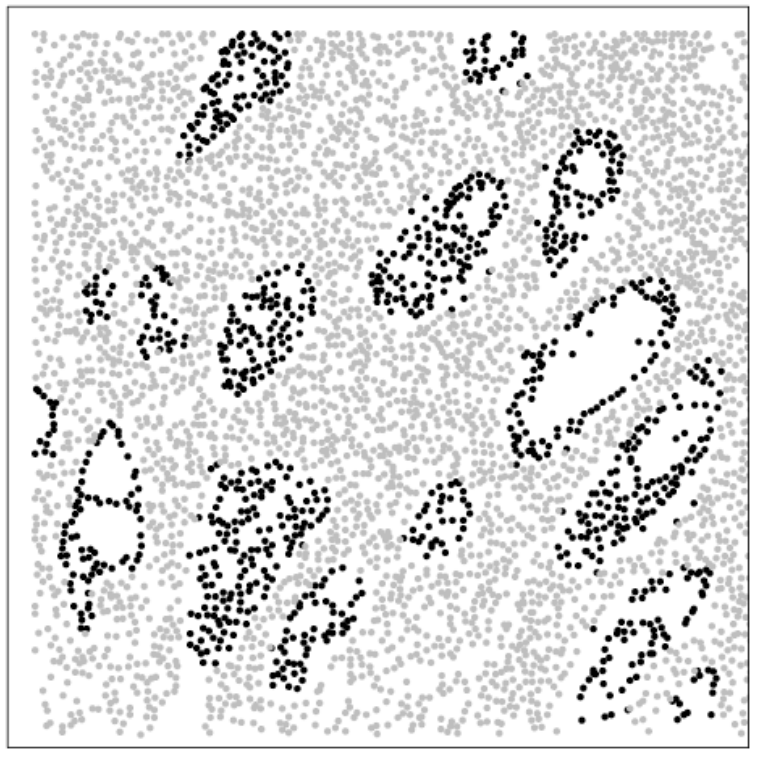}

\caption{Illustration of the image analysis of one patient's slide which enables the pinpointing of nuclei. Left: Composite immunofluorescence digital image showing Tumour (red), Stroma (green) and all nuclei (blue). Middle: Image analysis mask overlay from automatic machine learnt segmentation of the digital image. Tumour (purple), stroma (turquoise), necrosis (yellow). Right: Point pattern formed by the nuclei of the tumour (black) and stroma (grey) cells shown in the previous two images. }
\label{fig:slides}

\end{figure*}
Our data set pertains to 42 patients drawn from a wider data set of a pan-Scotland cohort of patients with CRC (see \citet{caie2014quantification} for a full description of our data set). In this subset of data, 23 patients died from CRC, and 19 survived. Each patient had up to 15 images from the invasive front of the tissue section processed by an automatic imaging algorithm described in \citet{caie2016next}. In summary, distinct regions in the digitised tissue section were first divided into four types through machine learning: (i) tumour, (ii) stroma, (iii) necrosis/lumen and (iv) no tissue. The positions of all cell nuclei were automatically identified (see \citet{caie2016next}) and exported by the image analysis software \citep{definiens}, see Figure \ref{fig:slides}. The centres of each nucleus provide us with our point patterns of tumour and stroma cells. In Section \ref{sec:voids.tumour} we consider tumour cells and stroma cells as separate void processes and seek to differentiate tissue sections based on patient survival. Then, in Section \ref{sec:thomas} we consider both tumour and stroma cells, fitting a Thomas process to each and using a graph theoretic approach to inform analysis of the spatial interaction between the cell types. In both these sections hierarchical bootstrap procedures were carried out following estimation of the associated parameters. A hierarchical bootstrap is required due to there being multiple images per patient. In particular, a patient is chosen at random, the data is subset, and then resampled. Then, the bootstrap is repeated a chosen number of times, in this case 1000. Finally, in Section \ref{sec:cv} we assess the predictive performance of each void and Thomas process through cross validation.

\subsection{Void processes for tumour and stroma cells}\label{sec:voids.tumour}

Recall that a void in the context of spatial point patterns is a region that contains no point where, given the general structure of the data, points may be expected. This section assumes that each of the separate point patterns formed by tumour and stroma nuclei are realisations of void processes, and so point patterns characterise the gaps in the  distribution of nuclei. In this way, the void process describing the tumour cell point pattern reflects the patterning of stroma cells and vice versa; note the void processes also reflect the less frequently occurring regions of necrosis/lumen and no tissue. 

\subsubsection{Results }\label{sec:results_voids}

Following estimation of $\theta = (R,D,\gamma)$ (for each of the stroma and tumour patterns) as per the methodology described above, a hierarchical bootstrap procedure (as detailed above), to account for patient effect, using 1000 simulations was carried out using the estimated parameters of interest for each patient group (i.e., those patients who were alive and those who died). Table \ref{tab:voids} contains the medians and bootstrapped quantities of the distributions of the parent densities, $\hat{D}$, and void radii, $\hat{R}$, pertaining to each point pattern (subscripts $t$ and $s$ refer to tumour and stroma patterns respectively). Here the superscripts  $d$ and $a$ refer to whether the patients were dead or alive at follow-up respectively. The results in Table \ref{tab:voids} reveal three findings in reference to the tumour void patterns: \begin{enumerate} [i)] \item There is no overlap between the patients who lived and the patients who died in the distribution of void densities; \item The values of the parent densities are lower for patients who lived than for patients who died; \item The estimated void radii pertaining to patients who lived and died are similar. \end{enumerate} When considering the stroma void patterns, there are no clear differences between patient  groups for density or radii.

\begin{table}[ht]
\centering
\caption{The medians and bootstrapped quantities (based on 1000 simulations) of the distributions of the parent densities and void radii for each of the point patterns formed by both tumour and stroma cells (subscript $t$ and $s$ respectively) of either patients who lived or died (superscripts $a$ and $d$ respectively) assuming a void process.}\label{tab:voids}
\begin{tabular}{rrrr}
  \hline
 & Median & 2.5\% Quantile & 97.5\%Quantile \\ 
  \hline
$\hat{D}_t^a$ & 8.6404 & 7.6931  &  9.4268 \\ 
  $\hat{D}_t^d$ &  15.4272 & 12.7820  & 18.2618\\ 
  $\hat{R}_t^a$ &  0.2142 & 0.1977  &  0.2305 \\ 
  $\hat{R}_t^d$ & 0.2152 &  0.1858   & 0.2417 \\ 
  $\hat{D}_s^a$ & 6.0304  & 5.3665  &  7.9776 \\ 
  $\hat{D}_s^d$ &  5.4828 & 5.0402  &  6.5505\\ 
  $\hat{R}_s^a$ &   0.2097 & 0.1929  &  0.2229 \\ 
  $\hat{R}_s^d$ &  0.2149 &  0.1976  & 0.2296 \\ 
   \hline
\end{tabular}
\end{table}

\subsection{Thomas processes for tumour and stroma cells }\label{sec:thomas}
In the context of histopathology, a Neyman Scott point process can be thought of as a process giving raise to the clustering of cells within the tissue. Therefore, we assume that the distribution of cells within each of the tumour and the stroma are assumed to be realisations of two different Neyman Scott point process. Observed cells are daughters and we seek to infer the unobserved parents. Parents thus represent some abstracted developmental process that led to the observed arrangement of daughter cells within the tissue.

\subsubsection{Framework and Results}\label{sec:cancer_results}

In order to fit Thomas processes to each pattern formed by tumour and stroma nuclei, we use the \texttt{R} package \texttt{nspp}\footnote{This package has been extended to incorporate parameter estimation for the Mat\'{e}rn process, and is available from \url{https://github.com/cmjt/nspp}} \citep{stevenson2015nspp}, to estimate $\boldsymbol{\theta}$ in each case. In order to fit the models it is required that we indicate the maximum distance we would expect a daughter to be from its parent, hence a reflection of the inter-cellular structure. This information is not supplied heuristically, but inferred from the use of a class cover catch digraph (CCCD) (details are given in Appendix B). The estimation of $\boldsymbol{\theta} = (D,\phi,\gamma)$  leads to the analytic derivation of daughter density, $\delta$.

As above, in order not to make any distributional assumptions, a hierarchical bootstrap (to account for patient effect) procedure is carried out using the estimated parameters of interest for each pattern. Table \ref{tab:thomas_cancer} contains the medians and bootstrapped quantities of the distributions of both the parent, $\hat{D}_\cdot^\cdot$, and daughter, $\hat{\delta}_\cdot^\cdot$, densities along with dispersion parameters---$\hat{\sigma}_\cdot^\cdot$, pertaining to each tumour and stroma point pattern. The results in Table \ref{tab:thomas_cancer} reveal 6 key findings: \begin{enumerate}[(i)] \item The values of both parent and daughter densities are higher for patients who lived than for patients who died in both stroma and tumour data; \item The differences in the distribution of both parent and daughter densities between patients who lived and patients who died are larger in the stroma data than in the tumour data; \item For tumour data, there is no overlap in the distributions between patients who lived and patients who died in the daughter densities, while parent density distributions overlap; \item For stroma data, there is no overlap in the distributions between patients who lived and patients who died in the parent or daughter densities; \item For dispersion, there is no overlap in the distributions between patients who lived and patients who died in the daughter densities for stroma data, while distributions overlap in the tumour data; \item For the stroma data, the dispersion is higher in patients who lived than patients who died.\end{enumerate}
For the distributions that did not overlap, the lowest value from one group is larger than the highest value in the other, thus indicating that the parameter estimates from a patient's slides are likely to be a good predictor for whether or not they survived. We formally assess the predictive power of this classifier into the following section.

\begin{table}[ht]
\centering
\caption{The medians and bootstrapped quantiles (based on 1000 simulations) of the distributions of both the daughter, $\hat{\delta}_\cdot^\cdot$, and parent, $\hat{D}_\cdot^\cdot$, densities along with dispersion parameters, $\hat{\sigma}_\cdot^\cdot$, for each of the point patterns formed by tumour and stroma nuclei---subscript $t$ and $s$ respectively---of either patients who lived or died---superscript $a$ and $d$ respectively, assuming a Thomas cluster process.}\label{tab:thomas_cancer}
\begin{tabular}{rrrr}
  \hline
 & Median & 2.5 \% Quantile & 97.5 \% Quantile \\ 
  \hline
   $\hat{\delta}_t^a$ & 2048.7671 &  1577.7571 & 2447.5904\\ 
  $\hat{\delta}_t^d$ & 1249.8016 &  962.0058 & 1550.2289 \\ 
  $\hat{\delta}_s^a$ & 3005.7470 & 1685.8138 & 4100.7619 \\ 
  $\hat{\delta}_s^d$ & 1057.6313 & 740.6736 & 1564.4209 \\ 
$\hat{D}_t^a$ & 91.6861 & 63.8080  & 273.3591\\ 
  $\hat{D}_t^d$ &  65.3319 & 47.2276 &  89.9136 \\ 
  $\hat{D}_s^a$ & 89.7333 & 39.1060 & 147.8126 \\ 
  $\hat{D}_s^d$ &  20.2366 & 9.2333  & 40.8440 \\ 
  $\hat{\sigma}_t^a$ & 0.0262&  0.0198  &  0.0339 \\ 
  $\hat{\sigma}_t^d$ & 0.0253 &  0.0215  &  0.0287 \\ 
  $\hat{\sigma}_s^a$ & 0.0335 &  0.0264  &  0.0404 \\ 
  $\hat{\sigma}_s^d$ & 0.0634 &  0.0463  &  0.0936\\
   \hline
\end{tabular}
\end{table}

%%%%%%%%%%%%%%%%%%%%%%%%%%%%%
\subsection{Cross validation}\label{sec:cv}

\begin{figure}
  \includegraphics[width=0.5\textwidth]{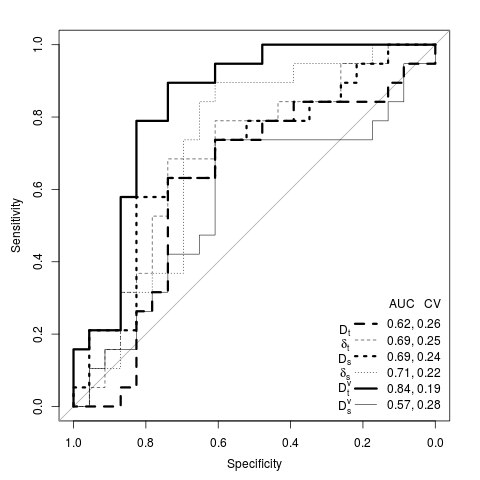}
  \caption{Plot of the ROC curves for each of the measures per patient. Legend includes the AUROC (area under the ROC curve) score for each curve and the CV (leave one out cross validation) score for each classifier (i.e., $D_\cdot$, $\delta_\cdot$, and $D^v_\cdot$ parent, daughter densities of a Thomas process and void density of a void process respectively for tumour and stroma patterns (subscripts $t$ and $s$ respectively))}
  \label{fig:ROC}
\end{figure}

We propose using the parameter estimates of the point processes discussed above to predict patients' outcomes. We assess the predictive power of each of these classifiers (estimates from each point process) using receiver operating characteristic (ROC) curves. Specifically, the means of each classifier were calculated per patient for use in a logistic regression (in predicting patient outcome). Figure \ref{fig:ROC} shows the ROC curves for each classifier along with the associated area under the ROC curve (AUROC) and leave one out cross validation (CV) scores. An AUROC score is a measure of classifier performance. To put our calculated AUROC scores in context, it should be noted that a score of 0.5 would indicate that our classifier does no better than random at distinguishing between patients who died or survived. In addition, a AUROC score of 1 would indicate perfect separation between each patient group (i.e., no overlap in distribution of our classifiers). The CV scores are an estimate of the test error for the logistic regression models using each classifier. From Figure \ref{fig:ROC}, notably, void density of the tumour pattern is the best predictor for patient outcome (AUROC of 0.84 and CV of 0.19). Further, when considering the AUROC scores, daughter density of the assumed Thomas process for the stroma patterns is considered the second best predictor. To asses goodness of fit of the assumed processes we compare the empty space function of the observed data to those patterns simulated with estimated parameter values. For further details see Appendix D where this concept is illustrated using the pattern shown in Figure \ref{fig:slides}.

\section{Discussion}\label{sec:disc}

Our void process analysis of gaps in tumour or stroma cells reflects the stroma or
tumour cell patterning, respectively as explained in Section \ref{sec:voids.tumour}. The tumour cell void
patterning is the most discriminatory and has the best predictive power: there is a
clear separation in parent densities (rows 1 and 2, Table \ref{tab:voids}); the best patient
outcome classifier is based on the gaps in tumour cells (AUROC=0.82, thick solid line of
Figure \ref{fig:ROC}). In contrast, the void process describing stroma cell gaps is the least
discriminatory and has the worst predictive power: there is no clear separation in
parent densities (rows 5 and 6, Table \ref{tab:voids}); the classifier based on stroma gaps has an
AUROC value approaching random (0.57, thin solid line of Figure \ref{fig:ROC}). Results \textbf{show that
  stroma patterning is a better indicator of patient outcome than tumour patterning}.

Notably, and in the context of this marked difference in performance, the radii of the
voids are very similar in both tumour and stroma patterns for all patients,
irrespective of whether they lived or died. This observation combined with the
result that parent density is lower in patients that lived than patients that died
means that there are \textbf{fewer voids in patients who lived and those voids are
  similar in size to those observed in patients who died}.

Our analysis using cluster point processes further supports the observation that stroma patterning is
more informative than tumour patterning. Parent and daughter densities are clearly
more distinct between patients who lived and patients who died for the stroma data
than for the tumour data, with densities being lower in patients who died. This
finding is also reflected in the cross-validation results, where a classifier based on
the Thomas process daughter density results in the second best AUROC (0.71, thin
dotted line of Figure \ref{fig:ROC}). We also observe a difference in the dispersion data: stroma
clusters are more dispersed in patients who lived than in patients who died; there is
no difference between patient groups in the tumour data. Thus \textbf{stroma cells
patterns have fewer, larger and more dispersed clusters in patients who lived
than in patients that died}.

These observations emerge from our unbiased data analysis and can be linked to
tumour-scale patterns in patients. High tumour budding is found in patients with an
infiltrative growth pattern: finger-like protrousions invading widely across the
stroma and thus forming large gaps between the cancer protrusions. Both tumour
budding and infiltrative growth pattern would be reflective of the tumour cell void
patterning described herein. In contrast, low tumour budding, or a pushing border
growth pattern, described as a solid tumour mass with little stroma existing
between cells, has been correlated with good outcome \citep{zlobec2009tumor}.
Although the literature reports these features to be significantly associated with
disease survival, the lack of consensus on quantification method, and observer
variability has led to their exclusion from clinical guidelines \citep{karamitopoulou2015tumour}. Our novel method provides a
standardised methodology that accurately describes and reports on tumour cell
distribution patterns which significantly correlate with patient outcome in CRC and
that negates observer variability.

Our method points to the importance of stromal features in predicting patient
outcome. The relevance of stroma patterns to patient outcome has also been
reported in breast cancer \citep{beck2011cognitive}, where the most differentiating measure was
a measure of fragmentation: patients with good outcomes had tissue that had larger
contiguous regions of stroma interspersed with larger epithelial regions; this
patterning is reflected in our nuclei-level analysis. More broadly, this work reflects a
growing interest in the use of analytical techniques more commonly associated with
ecology in recognition of the importance of both spatial structure and spatial
variations within that structure. For example, \citet{nawaz2015} observed that
both the abundance of immune cells and their spatial variation within the tumour
are important factors in patient outcome. Further, in a recent review of spatial
heterogeneity in cancers, \citet{heindl2015} set out the importance of such
ecological views of cellular patterns and the relevance of spatial statistics in
describing succinctly those patterns. Moreover they suggest the combined
exploration of spatial patterns of cells and cellular characteristics.

Here, we have presented a novel application of spatial statistics to capture the
spatial arrangement of cells through both void processes and clustered processes.
This work aligns with this growing recognition of the value of analytical methods
more typical in ecological contexts to understand the ecology of cancer. In the
future, we will build on our initial analysis by refining our treatment of individual
cells, stratifying individual cells further by biological and/ or physical
measurements to explore more complex questions that point at specific
mechanisms of disease progression.

\newpage 
\appendix
\section{The Palm intensity function}

This appendix derives the d-dimensional Palm intensity functions for both the void, and Mat\'{e}rn process discussed in this article. The d-dimensional Palm intensity for the modified Thomas process is derived by \citet{stevensonphd}. Our application considers only the 2-dimensional case; therefore, the article only provides details of the Palm intensities and likelihoods in 2-dimensions. This appendix generalises this to consider d-dimensions. Due to this we now consider the volume of hyper-spheres, and not the area of circles.
\subsection{A $d$-dimensional void point process}
\begin{figure}[htb]
  \centering
\includegraphics[width=0.5\textwidth]{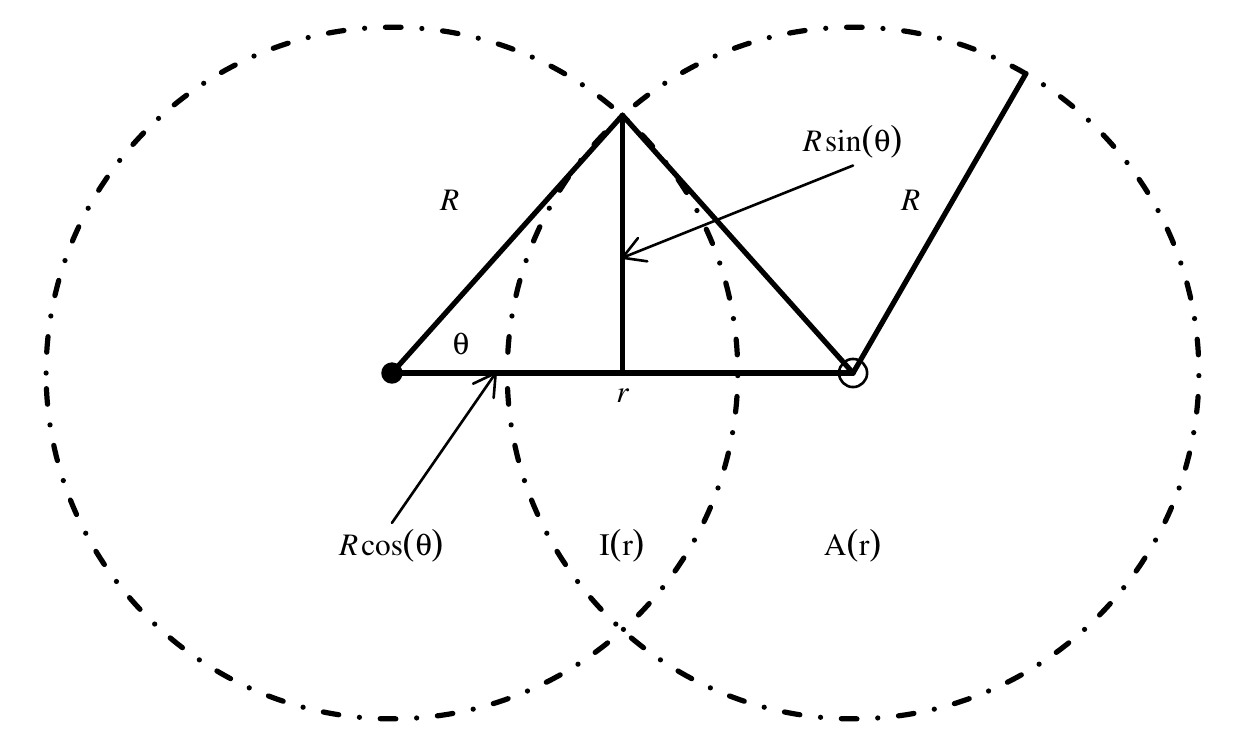}
\caption{\scriptsize{Figure showing the intersection of two spheres, in 2D, of radius $R$. The closed plotting character represents an observed daughter, the open plotting character represents a possible point. We know that there cannot exist a parent within distance $R$ of the observed daughter, therefore the intersection $I(r)$ is the area within which any possible point is ``safe''. To ascertain the intersection between two hyper-spheres of common radius $R$ the radius of the hyper-spherical caps, $R\,\text{sin}(\theta)$, and the height of the hyper-spherical caps $R\,\text{cos}(\theta)$ is required, where $\theta$ is the colatitude angle. Thus, the volume of intersection only depends on the radii of the hyper-spheres as well as the distance between their centers $r$, see Equation \ref{eq:intersection}. }}\label{fig:intersection}
  \end{figure}

The probability of a potential point being safe is related to the geometry of the intersection between hyper-spheres of common radius $R$ centered at the observed daughter (see Figure \ref{fig:simulated}, plot \textit{i)} encircled triangle) and a potential point, due to the expunging distance of parents (see Figure \ref{fig:simulated}, plot \textit{i)} dotted circles). This geometry is illustrated in Figure \ref{fig:intersection}.
  \begin{figure}[htb]
  \centering
  \includegraphics[width=0.5\textwidth]{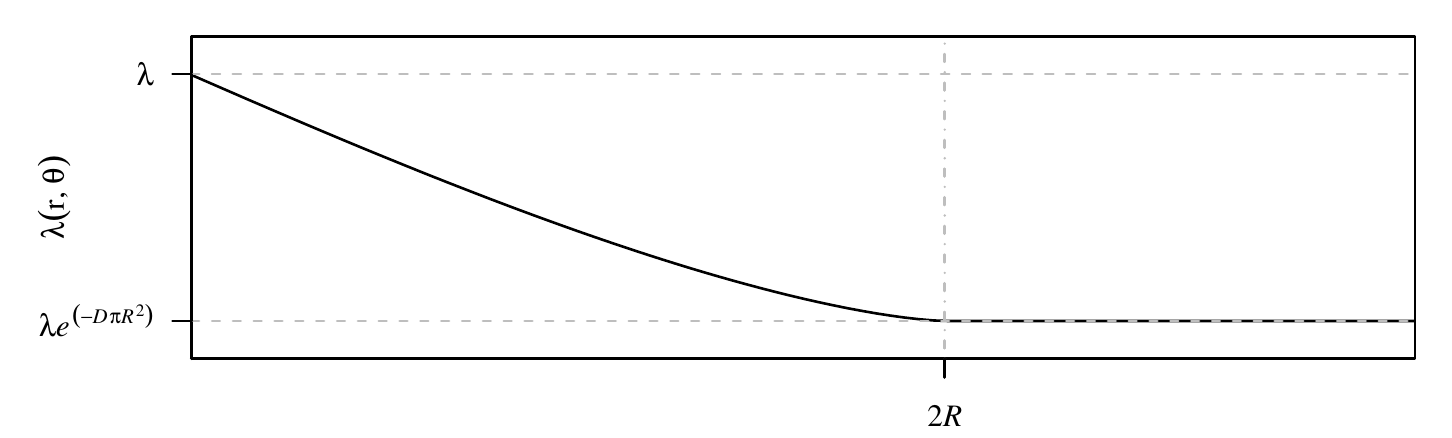}
  \caption{\scriptsize{The functional form of the Palm intensity for the void process in 2 dimensions. The horizontal asymptote is given by $\lambda\,\text{exp}(-D \pi R^2)$, which is the value that $\lambda_0(r)$ decays to for values of $r \geq 2\,R$. The Palm intensity at $r=0$ is simply $\lambda$, as for $r=0$ $g(r)=1$ in Equation \ref{eq:void.palm}, thus the exponential becomes $1$. At the value $r=2R$ the volume of intersection between the spheres encircling an observed daughter and a potential point of radius $R$ is zero, thus the contribution from the CDF of the Beta distribution to $\lambda(r;\theta)$ is zero.}}\label{fig:void.palm}
\end{figure}
The volume of intersection, $I(r)$ in Figure \ref{fig:intersection}, between two d-dimensional spheres with a common radius R is derived as in \citet{li2011concise}. To calculate $I(r)$, the integrand of  the volume of a $\text{d} -1$ sphere of radius $R\,\text{sin}(\boldsymbol{\theta})$ with height $R\,\text{cos}(\boldsymbol{\theta})$ is required. As the hyper-spheres are of common radius, $R$, $I(r)$ is simply just twice this volume. Hence;
\begin{equation}\label{eq:intersection}
  \begin{aligned}
    I^{\text{d}}(r) & = 2 \times \int_0^\phi v^{\text{d}-1}(R\text{sin}\boldsymbol{\theta})\,\text{d}R\text{cos}\boldsymbol{\theta},\\
    & =  2 \times \int_0^\phi v^{\text{d}-1}(R\text{sin}\boldsymbol{\theta})R\text{sin}\boldsymbol{\theta}\,\text{d}\boldsymbol{\theta},\\
    & =  2 \times \frac{\pi^{\frac{\text{d}-1}{2}}}{\Gamma(\frac{\text{d}-1}{2}+1)}\, R^\text{d}\int_o^\phi\text{sin}^{\text{d}}\boldsymbol{\theta} \,\text{d}\boldsymbol{\theta},\\
    & =   2 \times \frac{\pi^{\frac{\text{d}-1}{2}}}{\Gamma(\frac{\text{d}-1}{2}+1)}\, R^\text{d}\, \text{J}_d(\phi),\\
    & = \frac{\pi^{\frac{\text{d}-1}{2}}}{\Gamma(\frac{\text{d}-1}{2}+1)}\, R^{\text{d}}\,B\left(\frac{\text{d}+1}{2},\frac {1}{2}\right)\,\text{I}\left(\text{sin}^2\phi;\frac{\text{d}+1}{2},\frac{1}{2}\right),\\
    & = v^{\text{d}}(R)\,\text{I}\left(\text{sin}^2\phi;\frac{\text{d}+1}{2},\frac{1}{2}\right),\\
  & = v^{\text{d}}(R)\,\text{I}\left(1-\left(\frac{r}{2R}\,\right)^2;\frac{\text{d}+1}{2},\frac{1}{2}\right),\\
  \end{aligned}
  \end{equation}
noting that $B(a,b)=\frac{\Gamma(a)\,\Gamma(b)}{\Gamma(a+b)}$, $\Gamma\left(\frac{1}{2}\right) = \sqrt{\pi}$, and that $(R\,\text{cos}\boldsymbol{\theta})^2 + (R\,\text{sin}\boldsymbol{\theta})^2 = R^2 \rightarrow (\text{sin}\,\boldsymbol{\theta})^2 = 1 - (\text{cos}\boldsymbol{\theta})^2$ and using the cosine rule leads to  $\text{cos}\,\boldsymbol{\theta} = \frac{r^2+R^2-R^2}{2\,r\,R} = \frac{r}{2\,R}$. Also $\text{I}(z;a,b)=\frac{B(z;a,b)}{B(a,b)}$ is the regularised Beta function. 

The Palm intensity of the void process then becomes;
\begin{equation}\label{eq:void.palm.2}
  \begin{aligned}
\lambda_0(r) & =\lambda\,\text{exp}\left(-D\,v^{\text{d}}(R)\left[1-\text{I}\left(1-\left(\frac{r}{2R}\,\right)^2;\frac{\text{d}+1}{2},\frac{1}{2}\right)\right]\right),\\
& = \lambda\,\text{exp}\left(-D\,v^{\text{d}}(R)\left[1-\text{F}_{g(r)}\left(\frac{\text{d}+1}{2},\frac{1}{2}\right)\right]\right),
\end{aligned}
\end{equation}
  where $g(r)=1-\left(\frac{r}{2R}\,\right)^2$, and $\text{F}_{g(r)}(\cdot,\cdot)$ is the CDF of the Beta distribution. Thus; when $r=0$ $\Rightarrow$ $g(r)=1$ $\Rightarrow$ $\text{F}_1(\cdot,\cdot)=1$ $\Rightarrow$ $\lambda_0(0)=\lambda$, in addition when $r=2R$ $\Rightarrow$ $g(r)=0$ $\Rightarrow$ $\text{F}_0(\cdot,\cdot)=0$ $\Rightarrow$ $\lambda_0(0)=\lambda\,\text{exp}(-D\,v_\text{d}(R))$, due to the properties of the CDF. The functional form of this Palm intensity is shown in Figure \ref{fig:void.palm}. Substituting d$=2$ into Equation \eqref{eq:void.palm.2} would lead to the Palm intensity given in Equation \eqref{eq:void.palm}.
  
\subsection{A $d$-dimensional  Mat\'{e}rn point process}
\begin{figure}[htb]
  \centering
  \includegraphics[width=0.5\textwidth]{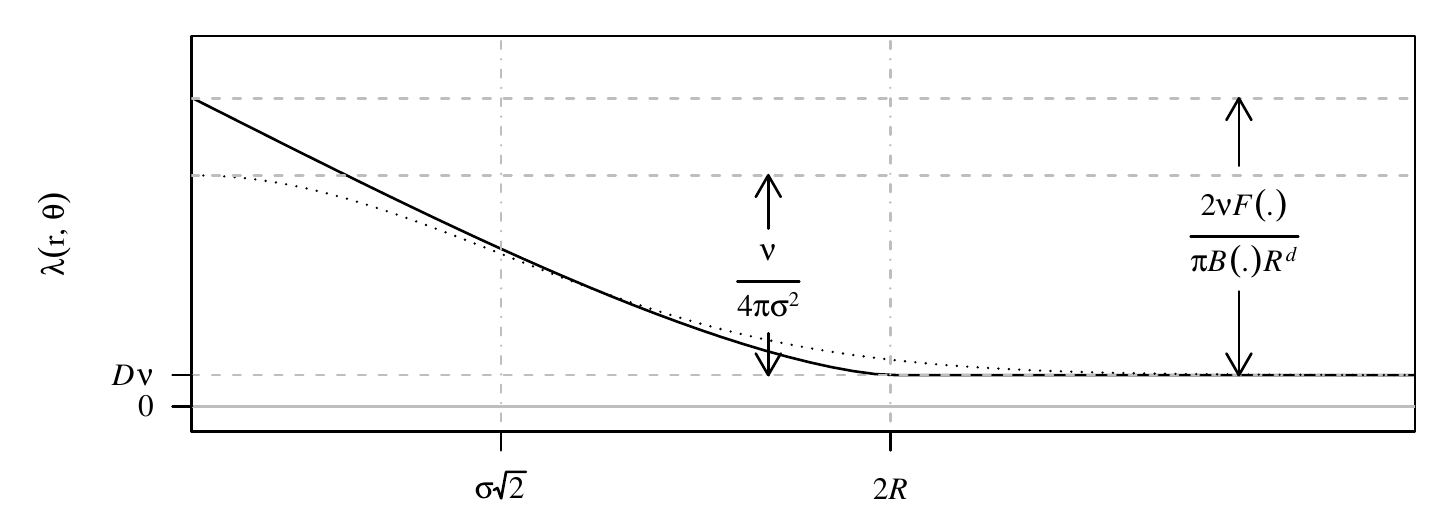}
  \caption{\scriptsize{The functional form of the Palm intensity for both the Mat\'{e}rn (solid curve) and Thomas (dotted curve) variants of the Neyman Scott point process in 2 dimensions. The horizontal asymptote is given by $D\,\nu$ for both processes. The difference between the horizontal asymptote and the $y-intercept$ is given by $\nu/(4\,\pi\,\sigma^2)$ for the Thomas process, and $2\,\nu\,F(\cdot)/(\pi\,B(\cdot)\,R^2)$ for the Mat\'{e}rn process. Here $F(\cdot) = {}_2F_1\left( \frac {1} {2},\frac {1} {2}- \frac {\text{d}} {2}, \frac {3} {2},1\right)$ denotes the hyper-geometric function, and $B(\cdot) = B(\frac{3}{2},\frac{1}{2})$ denotes the beta function. The point of inflection of the Gaussian term for the Thomas process is given by $\sigma\,\sqrt(2)$, for the Mat\'{e}rn process $2\,R$ is the point at which the Palm intensity decays to the horizontal asymptote.}}\label{fig:palm}
\end{figure}

Daughters of a Mat\'{e}rn process are uniformally distributed around their parents. The parameter $\gamma$ in Equation \ref{eq:palm.intensity} refers to the radius of the sphere centered at a selected parent outwith which we do not observed sired daughters. Figure \ref{fig:simulated} illustrates the two types of Neyman Scott point processes simulated with the same value of $\gamma$. Thus, for the same value of $\gamma$ clearly the Palm intensity function for the Mat\'{e}rn process is initially much higher, and decays at a much faster rate to the horizontal asymptote, $D\,\nu$. As illustrated in Figure \ref{fig:palm}, letting $\gamma = R$, $\lambda(r;\boldsymbol{\theta})$ is a continuous piece-wise monotonic function of two sub-domains, $[0,2R]$ and $[2R,\infty)$. The common endpoint of the sub-domains, $2R$, relates to the structure of the Mat\'{e}rn process, that is, the distance between two sibling daughters cannot be more than the diameter, $2R$, of a sphere centered at an unobserved parent away from one another. Thus, the probability of observing a sibling at a distance $r$ from an arbitrarily chosen daughter pertains to the intersection of the hyper-spheres centered at these points, $b(x,R) \cap b(y,R), {x \neq y} \in N$, where $N$ is the point pattern. Clearly when the distance between these points, $r \geq 2R$ then $b(x,R) \cap b(y,R) = 0$.

The d-dimensional version of Equation \ref{eq:mat.dist} is given by,
  
\begin{equation}\label{eq:mat.dist.2}
  \begin{aligned}
    f_y^\text{d}(r;R) &= \frac{2\,\text{d}}{B(\frac {\text{d}} {2} + \frac {1} {2},\frac{1}{2})}\, \frac{r^{\text{d}-1}}{R^{\text{d}+1}}\, \left[ {}_2F_1\left( \frac {1} {2},\frac {1} {2}- \frac {\text{d}} {2}, \frac {3} {2},1\right)\,R - {}_2F_1\left(\frac {1} {2},\frac {1} {2} - \frac {\text{d}} {2}, \frac {3} {2}, \frac {r^2}{4\,R^2}\right)\,\frac {r} {2} \right], \\
    & = \frac {2 \, \text{d} \,r^{\text{d} -1}\, \int^R_{\frac{r}{2}} (R^2 - x^2)^{\frac {\text{d} -1}{2}} dx} {B(\frac {\text{d}} {2} + \frac {1} {2},\frac{1}{2})\, R^{2\,\text{d}}}.
    \end{aligned}
  \end{equation}
Here $B(\cdot,\cdot)$ denotes the beta function, and ${}_2F_1(\cdot,\cdot,\cdot,\cdot)$ the hyper-geometric function.
  
Below we show how this PDF reduces in $d=2$ and $d=3$ to forms equivalent to the PDFs of the distances between two randomly selected siblings given by \citet[p.~376]{illian2008statistical} in the respective dimensions.

 $\bullet$ for $\text{d} =2 $
  It should be noted that,
  \begin{equation*}
    \begin{aligned}
      \int^R_{\frac{r}{2}} (R^2 - x^2)^{\frac {\text{d} -1}{2}} dx & = \int^R_{\frac{r}{2}} (R^2 - x^2)^{\frac {1}{2}} dx\\
          & = \frac{1}{8} \left(4\,R^2\,\text{sec}^{-1}\left(\frac{2\,R}{r}\right) - r\,\sqrt{4\,R^2 - r^2}\right).
    \end{aligned}
  \end{equation*}
  therefore,
  \begin{equation*}
    \begin{aligned}
      \frac {2 \, \text{d} \,r^{\text{d} -1}\, \int^R_{\frac{r}{2}} (R^2 - x^2)^{\frac {\text{d} -1}{2}} dx} {B(\frac {\text{d}} {2} + \frac {1} {2},\frac{1}{2})\, R^{2\,\text{d}}} & = \frac {4 \,r\, \int^R_{\frac{r}{2}} (R^2 - x^2)^{\frac {1}{2}} dx} {B(\frac {\text{3}} {2},\frac{1}{2})\, R^{4}}\\
& =  \frac{r\,\Gamma(2)}{\frac{1}{2} \sqrt{\pi}^2\,R^4\,2} \left(4\,R^2\,\text{sec}^{-1}\left(\frac{2\,R}{r}\right) - r\,\sqrt{4\,R^2 - r^2}\right)\\
& = \frac {r}{\pi\,R^4}\,\left(4\,R^2\,\text{sec}^{-1}\left(\frac{2\,R}{r}\right) - r\,\sqrt{4\,R^2 - r^2}\right)\\
& = \frac {r}{\pi\,R^4}\,\left(4\,R^2\,\text{cos}^{-1}\left(\frac{r}{2\,R}\right) - r\,\sqrt{4\,R^2 - r^2}\right),\\
    \end{aligned}
  \end{equation*}
noting that $\text{sec}^{-1}(x) = \text{cos}^{-1}\left(\frac{1}{x}\right)$, and as $B(\frac {\text{3}} {2},\frac{1}{2}) = \frac{\Gamma(\frac{3}{2})\,\Gamma(\frac{1}{2})}{\Gamma(2)}$, recalling that $\Gamma(\frac{3}{2}) = \frac{1}{2}\sqrt{\pi}$, $\Gamma(\frac{1}{2}) = \sqrt{\pi}$, and $\Gamma(2) = 1$.

$\bullet$for $\text{d} =3$
 It should be noted that,
  \begin{equation*}
    \begin{aligned}
      \int^R_{\frac{r}{2}} (R^2 - x^2)^{\frac {\text{d} -1}{2}} dx & = \int^R_{\frac{r}{2}} (R^2 - x^2) dx\\
          & = \frac{1}{24} \, \left(r - 2\,R\right)^2\,\left(r + 4\,R\right).
    \end{aligned}
  \end{equation*}
  therefore,
  \begin{equation*}
    \begin{aligned}
      \frac {2 \, \text{d} \,r^{\text{d} -1}\, \int^R_{\frac{r}{2}} (R^2 - x^2)^{\frac {\text{d} -1}{2}} dx} {B(\frac {\text{d}} {2} + \frac {1} {2},\frac{1}{2})\, R^{2\,\text{d}}} & = \frac {6 \,r^{2}\, \int^R_{\frac{r}{2}} (R^2 - x^2)dx} {B(2,\frac{1}{2})\, R^{6}}\\
& =  \frac {6 \,r^{2}} {24 \, B(2,\frac{1}{2})\, R^{6}}\, \left( r - 2\,R\right)^2\,\left(r + 4\,R \right),\\
& =  \frac{r^2\, \Gamma(\frac{5}{2})}{4\, R^6\, \Gamma(2)\,\Gamma(\frac{1}{2})}    \, \left( r - 2\,R\right)^2\,\left(r + 4\,R \right)\\
& = \frac{r^2\, \frac{3}{4}\,\sqrt{\pi}}{4\, R^6\, \sqrt{\pi}}    \, \left( r - 2\,R\right)^2\,\left(r + 4\,R \right)\\
& = \frac{3 \,r^2}{16\, R^6}    \, \left( r - 2\,R\right)^2\,\left(r + 4\,R \right)\\
& = \frac{3\,r^2}{16\,R^6}\left(R-\frac{r}{2}\right)^2\left(2\,R + \frac{r}{2}\right).\\
    \end{aligned}
  \end{equation*}
noting that $\Gamma(\frac{5}{2}) = \frac{3}{4}\sqrt{\pi}$, $\Gamma(\frac{1}{2}) = \sqrt{\pi}$, and $\Gamma(2) = 1$.

Upon substitution of the PDF, given by Equation \ref{eq:mat.dist.2}, into the Palm intensity function, given by Equation \ref{eq:palm.intensity}, as in the case of the modified Thomas process above simplifications occur which circumvent numerical instability in $\lambda(r;\boldsymbol{\theta})$ at $r=0$ as both the numerator and denominator in the second term contain the term $r^{d-1}$. Thus, 

 \begin{equation}\label{eq:palm.intensity.mat}
  \begin{aligned}
  \lambda(r;\boldsymbol{\theta}) & = D\,E_c(\phi) + \frac{[E_s(\phi)-1]\,f_y^\text{d}(r;\gamma)}{s^\text{d}(r)}, \\
   &=  D\,E_c(\phi) + \frac{[E_s(\phi)-1]\,2\,\text{d}}{B(\frac {\text{d}} {2} + \frac {1} {2},\frac{1}{2})}\, \frac{r^{\text{d}-1}}{R^{\text{d}+1}}\, \frac{\Gamma(\frac{\text{d}}{2}+1)}{\text{d}\,\pi^{d/2}\,r^{\text{d}-1}} \\
  & \times \left[ {}_2F_1\left( \frac {1} {2},\frac {1} {2}- \frac {\text{d}} {2}, \frac {3} {2},1\right)\,R  -  {}_2F_1\left(\frac {1} {2},\frac {1} {2} - \frac {\text{d}} {2}, \frac {3} {2}, \frac {r^2}{4\,R^2}\right)\,\frac {r} {2} \right],\\
 &=  D\,E_c(\phi) + \frac{2\,[E_s(\phi)-1]}{B(\frac {\text{d}} {2} + \frac {1} {2},\frac{1}{2})R^{\text{d}+1}}\, \frac{\Gamma(\frac{\text{d}}{2}+1)}{\pi^{d/2}} \\
  & \times \left[ {}_2F_1\left( \frac {1} {2},\frac {1} {2}- \frac {\text{d}} {2}, \frac {3} {2},1\right)\,R  -  {}_2F_1\left(\frac {1} {2},\frac {1} {2} - \frac {\text{d}} {2}, \frac {3} {2}, \frac {r^2}{4\,R^2}\right)\,\frac {r} {2} \right],\\
&=  D\,E_c(\phi) + \frac{2\,[E_s(\phi)-1]}{R^{\text{d}+1}\,\pi^{\frac{\text{d}}{2}}}\frac{\Gamma(\frac{\text{d}}{2}+1)^2}{\Gamma(\frac{\text{d}}{2}+\frac{1}{2})\,\Gamma(\frac{1}{2})} \\
  & \times \left[ {}_2F_1\left( \frac {1} {2},\frac {1} {2}- \frac {\text{d}} {2}, \frac {3} {2},1\right)\,R  -  {}_2F_1\left(\frac {1} {2},\frac {1} {2} - \frac {\text{d}} {2}, \frac {3} {2}, \frac {r^2}{4\,R^2}\right)\,\frac {r} {2} \right],\\
&=  D\,E_c(\phi) + \frac{2\,[E_s(\phi)-1]}{R^{\text{d}+1}\,\pi^{\frac{1}{2}(\text{d}+1)}}\frac{\Gamma(\frac{\text{d}}{2}+1)^2}{\Gamma(\frac{\text{d}}{2}+\frac{1}{2})} \\
  & \times \left[ {}_2F_1\left( \frac {1} {2},\frac {1} {2}- \frac {\text{d}} {2}, \frac {3} {2},1\right)\,R  -  {}_2F_1\left(\frac {1} {2},\frac {1} {2} - \frac {\text{d}} {2}, \frac {3} {2}, \frac {r^2}{4\,R^2}\right)\,\frac {r} {2} \right].\\
  \end{aligned}
  \end{equation}
 noting that $B(x,y)=\frac{\Gamma(x)\,\Gamma(y)}{\Gamma(x+y)}$, and $\Gamma(\frac{1}{2})=\sqrt{\pi}$.

%% \subsection{Parameter estimation limitations}

%% It should be noted that the set of parameters pertaining to each of the void---$\theta = (D,R,\gamma)$---and cluster---$\theta = (D,\phi,\gamma)$---processes giving raise to a certain Palm likelihood are doubtfully unique. That is, there may exist other sets of parameters, considering the void---$\theta^* = (D^*,R^*,\gamma^*)$---and cluster---$\theta^* = (D^*,\phi^*,\gamma^*)$---processes respectively, which gives rise to the same Palm likelihood as the sets $\theta$ above. Specifically, in the case of both the 2D void and  Mat\'{e}rn cluster process, considering the Palm intensity functions (Appendix A, Figures \ref{fig:void.palm} and \ref{fig:palm}) an inverse proportional relationship exists between the parameters $D$ and $R^2$. This is due to the geometry of the assumed spatial structure of the process. In each case comparison of the fitted and the empirical Palm intensity will aid in a somewhat heuristic way in assessing model fit.

\section{Class Cover Catch Digraphs}

Due to the structure of our data comprising two membership groups of tumour and stroma, and in order to estimate $\boldsymbol{\theta}$, information with regards to the heterogeneous nature of the tissue structure is required. This information can inform the starting value for the fitting procedure to facilitate convergence. Using a technique derived from graph theory this may be thought of as a class cover problem \citep{cannon1998approximate}, whereby an algorithmic approach is taken in order to form two disjoint groups. Data random digraphs have often been used in the context of spatial
pattern recognition \citep{marchette2005random}, and a subtype of these
graphs termed a class cover catch digraph (CCCD) \citep{priebe2001distribution} has been used to
classify data in multiple dimensions \citep{devinney2006new}. Generally, proximity graphs offer
invaluable information as to the neighbourhood structure and
interaction among two or more classes. 
However, in this context restricting the pattern such that---what may be thought of as---cluster density of one class of point is forced by the locations of the second class makes an unnecessary assumption about interactions between classes. Thus, rather than using characteristics of the CCCD to classify the pattern, we use calculated summaries to inform the likely maximum distance we would expect a daughter to be from its parent in order to aid optimisation. Note, assuming that each class of point is a realisation of a Thomas process does not suppose for instance that the spatial region occupied by a cluster of tumour cells cannot also be occupied by stroma cells.  

Data random digraphs have often been used in the context of spatial
pattern recognition \citep{marchette2005random}, a sub-type of these
graphs named Class Cover Catch Digraph (CCCD)s \citep{priebe2001distribution} have been used to
classify data in multiple dimensions \citep{devinney2006new}. Generally proximity graphs offer
invaluable information as to the neighbourhood structure and
interaction among two or more classes, however the choice of graph can
often bias the spatial analysis \citep{rajala2010spatgraphs}.

A simple graph is a pair of sets $(V,E)$, where E is the set
of edges connecting the set of $V={v_i, ..., v_n}$ verticies, a
directed graph (digraph), is such where the edges are ordered
and as such are denoted by $(V,D)$. Consider the situation  where
there are observations from two classes $X ={x_1, ..., x_n}, Y =
{y_1, ... , y_m} \in \mathbf{R}^q$, the \textit{class cover} problem
aims to find the smallest collection of ``spheres'' centered at
observations in $X$ such that every observation in $X$ is in at least
one of the ``spheres'' and no observation in $Y$ is in any
``sphere''. If we now consider a collections of sets ${B_1, B_2,
...}$ with associated ``base'' points ${t_1, t_2, ...}$ a
catch digraph $(V,D)$ has a directed edge from $v_i$ to $v_j$
if and only if $t_j \in B_i$. Therefore for any sets $X,Y \in
\mathbf{R}^q$ the class cover catch digraph (CCCD) to be the
\textit{catch digraph}  formed by base points $x_i \in X$ and the
associates sets $B_i = {z \in \mathbf{R}^q : d(x_i,z) < d(x_i,Y)}$.

One may na\"{i}vely suppose that the sets ${B_1, B_2,...}$ of a CCCD and the parent density, $D$, relating to one class of point proffer the same information. However, here it is conceptualised that these parents points are in fact abstract concepts pertaining to the process of cell division, growth, death and interaction within the micro-environment of the tumour tissue. Thus, it is not assumed that the parents of each class, which would be the base points $x_i \in X$ and $y_i \in Y$ in the CCCD methodology, are observed. Thence, a robust measure which represents the latent process proffering the spatial distribution of points, taking into account the interactions between the classes is required, hence, the CCCD--NSPP approach is employed.

\section{R packages}
All code used in the application of the methodology discussed is supplied in the supplementary material. Please note that both packages referred to below are still undergoing development at the time of submission, hence documentation may be sparse.
\subsection*{The \texttt{gapski} \texttt{R} package}
  The \texttt{R} package \texttt{gapski} is not yet available on \texttt{CRAN}, but is available from \url{https://github.com/cmjt/gapski}. This implements the methods pertaining to void processes described in this article. The main functionality of this package is the fitting of a void process to point pattern data, which is carried out through the function \texttt{fit.gap()}. Parameter estimation of $\theta = (R,D,\lambda)$ is carried out using the following code;\\
  \begin{lstlisting}[language=R]
 fit.gap(points = points,lims = lims,trunc = trunc,
         D.sv = D.sv, D.bounds = D.bounds)
   \end{lstlisting}
Further details can be seen in the documentation provided.
\subsection*{Extensions to the \texttt{nspp} \texttt{R} package}
Parameter estimation of a NSPP, following the methodology described above, is carried out  through using the \texttt{R} package \texttt{nspp} package, not yet available on \texttt{CRAN} but from \url{https://github.com/cmjt/nspp}. Please note that this is an extension of the  \texttt{nspp} package---version 1.0---of \citet{stevenson2015nspp}. Where the sole contribution by the authors was the derivation and hence inclusion of the Palm intensity of a Mat\'{e}rn process enabling a approximate-likelihood approach to model fitting to be taken.
Parameter estimation of $\theta = (R,\phi,\gamma)$ can be carried out using the following code;\\
\begin{lstlisting}[language=R]
fit.ns(points = points, lims = lims, R = R,
       dispersion = dispersion)
                   \end{lstlisting}
where the argument \texttt{dispersion} is either \texttt{"gaussian"} or \texttt{"uniform"}, hence fitting either a Thomas or Mat\'{e}rn process respectively. Further details can be seen in the documentation provided.

\section{Assessing Model fit using the empty space function}

\begin{figure}[htb]
  \centering
\includegraphics[width = 0.5\textwidth]{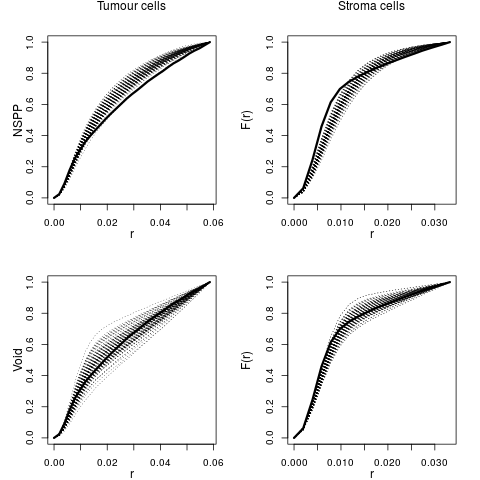}
\caption{\scriptsize{Figure showing for one patient's slide (Figure \ref{fig:slides}) the fitted empty space function, F(r), by the solid line, and those for 100 simulated patterns (simulated using the estimated parameter values for each process), dotted lines.}}\label{fig:model.check}
  \end{figure}

In order to asses the suitability of the considered point processes, of which the cell nuclei were assumed to be a realisation, one might wish to  compare the spatial patterning of the observed data and data simulated with the estimated parameter values. To do this we use the empty space function, and compare this for the observed data to those for patterns simulated with the respective estimated parameter values. Figure \ref{fig:model.check} shows the empty space functions for each process (i.e., modified Thomas process top row, and void process bottom row) for the slide shown in Figure \ref{fig:slides}. Each solid line represents the empty space function for the fitted model; the dashed lines are the empty space functions estimated for patterns simulated with the respective estimated parameters. From Figure \ref{fig:model.check} we can see that the considered NSPP closely approximates the spatial patterning of cells, and that the void process is well fitted to the observed data.

\newpage

\bibliography{bib} 
\bibliographystyle{imsart-nameyear}

\end{document}